
 \magnification\magstep 1
\vsize=24 true cm
\hsize=16 true cm
\baselineskip=14pt
\parindent=1 true cm
\parskip=0.3 true cm
\def\oppropto{\mathop{\propto}}

\vglue 4 true cm

\centerline{ \bf DIFFUSIVE TRANSPORT IN A ONE DIMENSIONAL  }
\centerline{ \bf}
\centerline{ \bf  DISORDERED POTENTIAL INVOLVING CORRELATIONS}

\vskip 2 true cm

\centerline{   C\'ecile Monthus \footnote{$^{(1)}$}{monthus@ipncls.in2p3.fr} }

\vskip 1 true cm
\centerline{ {\it Division de Physique Th\'eorique} \footnote{$^{(2)}$}{Unit\'e
de recherche des Universit\'es Paris 11 et Paris 6 associ\'ee au CNRS} {\it ,
IPN
B\^at. 100, 91406 ORSAY, France} }
\centerline{ \bf}
\centerline{ and}
\centerline{ \bf}
\centerline{\it L.P.T.P.E., Universit\'e Paris 6, 4 Place Jussieu, 75252 PARIS,
France}

\vskip 2 true cm

\noindent { \bf Abstract}

This article deals with transport properties of one dimensional Brownian
diffusion
under the influence of a correlated quenched random force, distributed as a
two-level
Poisson process. We find in particular that large time scaling laws of the
position
of the Brownian particle are analogous to the uncorrelated case. We discuss
also the
 probability distribution of the stationary flux going
through a sample between two prescribed concentrations, which differs
significantly from
the uncorrelated case.


\vfill

\line {IPNO/TH 95-19  \hfill }

\eject

Disorder significantly influences diffusive transport phenomena
[1] [2] [3]. Indeed, strong enough disorder does not induce
a simple renormalization for transport coefficients of the corresponding pure
system, but usually generates "anomalous diffusion" effects.
The case of spatially uncorrelated disordered media is now pretty well
understood.
The effects of spatial correlations on the dynamics
have first been discussed for the directed random walk problem [4].
More recently, the response to an external applied force has been studied
for the one dimensional case of a Brownian particle diffusing under the
influence of a quenched random force $\{F(x)\}$ [5].
For a sample characterized by some particular realization of the
stochastic process $\{F(x)\}$, the diffusion is
 defined by the following
 Fokker-Planck equation for the probability density $P(x,t \  \vert x_0, 0)$
$$
{{\partial P} \over {\partial t }} = D_0 {\partial \over {\partial x}}
\left( {\partial P\over {\partial x}} - \beta F(x) P \right)
\eqno (1)
$$
In this article, we analyse in details transport properties of this model
in the case where $\{F(x)\}$ is
 distributed as a two-level Poisson process.

Let us define more precisely the model of disorder we consider
and some of its properties.
We assume that the quenched random force $F(x)$
takes alternatively a positive value $\phi_0>0$ and
a negative value $-\phi_1<0$ on intervals whose lengths are independent random
variables
distributed according to the following probability densities respectively
$$\left\{\matrix{
f_0(l)= \theta(l) \ n_0 \ e^{-n_0l}\hfill\cr
\noalign{\medskip}
f_1(h)= \theta(h) \ n_1 \ e^{-n_1h}\hfill\cr
}\right.
\eqno(2)
$$
The parameters ${1 \over n_0}$ and ${1 \over n_1}$ are respectively the mean
length
of intervals $\{F(x)=\phi_0\}$ and the mean length of intervals
$\{F(x)=-\phi_1\}$.
This choice of exponential distributions for $f_0$ and $f_1$ is in fact the
only
one that makes the process
$\{F(x)\}$ Markovian. This property enables us to write differential
equations for the probability $p_0(x)$ to have $F(x)=\phi_0$ and the
probability $p_1(x)=1-p_0(x)$ to have $F(x)=-\phi_1$.
$$\left\{\matrix{
\displaystyle {\partial  p_0\over \partial
x}=-n_0p_0+n_1p_1=n_1-(n_0+n_1)p_0\hfill\cr
\noalign{\medskip}
 \displaystyle {\partial p_1\over \partial
x}=-n_1p_1+n_0p_0=n_0-(n_0+n_1)p_1\hfill\cr
}\right.
\eqno(3)
$$
The mean value $F_0$ and the two point correlation function $G(x)$ of the
process $\{F(x)\}$ read
$$\left\{\matrix{
F_0\equiv <\phi>=\phi_0 \displaystyle{n_1\over n_0+n_1}-\phi_1{n_0\over
n_0+n_1}\hfill\cr
\noalign{\medskip}
G(x)\equiv <\phi(x)\phi(0)>-<\phi>^2= \displaystyle{n_0n_1\over(n_0+n_1)^2}
(\phi_0+\phi_1)^2e^{-(n_0+n_1)\vert x\vert}\hfill\cr
}\right.
\eqno(4)
$$

The corresponding random potential $U(x) = -\int^x F(y) dy$
seen by the Brownian particle presents an alternance of positive and negative
slopes
of random lengths as sketched on Figure 1.

The fundamental random variable associated with classical diffusion under the
action
of a quenched random force presenting a stricly
positive mean $<F(x)> \equiv F_0 >0$, is the exponential functional of the
random potential $U(x)$
$$
\tau_{\infty} = \int_0^{\infty} dx \ e^{ \displaystyle \beta U(x) }
 = \int_0^{\infty} dx \ e^{ \displaystyle - \beta \int_0^x F(y) dy}
\eqno (5)
$$
Indeed,
the probability distribution of the functional $\tau_{\infty}$
determines the large time anomalous behaviour of the Brownian particle position
[3].
 In particular, the velocity defined for each sample as
$$
V = \lim_{t \to \infty} {d \over dt} \int_{- \infty}^{+ \infty}
 dx \ x P(x,t \  \vert x_0, 0)
\eqno (6)
$$
is a self-averaging quantity [6] (for any homogenous random potential
presenting
only short-range correlations) inversely proportional to the first moment of
 $\tau_{\infty}$ [3]
$$
V= {D_0 \over {<\tau_{\infty} >}}
\eqno (7)
$$
When the quenched random force is distributed with the Gaussian measure
$$
{\cal D} F(x) \ e^{ \displaystyle -{1 \over {2 \sigma} } \int dx \
[F(x)-F_0]^2}
\eqno (8)
$$
the probability distribution ${\cal P} _{\infty} (\tau)$ of the functional
$\tau_{\infty}$ reads in terms of $ \alpha = {{\sigma \beta^2} \over 2}$
and $\mu = {2 F_0 \over {\beta \sigma}} >0$ [3]
$$
{\cal P} _{\infty} (\tau) = {\alpha \over \Gamma(\mu)} \left( {1 \over \alpha
\tau}
\right)^{1+\mu} \ e^{ \displaystyle - {1 \over \alpha \tau}}
\ \ \oppropto_{ \tau \to \infty} \ \ {1 \over \tau^{1+\mu}}
\eqno (9)
$$
The algebraic decay at large $\tau$ explains all the dynamical phases
transitions
between different anomalous behaviours known for this model [3]. In particular,
Eq (7) implies that the value $\mu = 1$ separates a phase of vanishing
velocity $V=0$ for $0<\mu<1$, and a phase of finite velocity $V>0$ for $\mu
>1$.
Another interesting physical quantity is
the stationnary current $J_N$ which goes through a disordered
sample of length $N$ between a fixed concentration $P_0$ and a trap
described by the boundary condition $P_N=0$ [7] [8].
In the limit $N \to \infty$,
the stationary flux $J_{\infty} $ is simply a random variable inversely
proportional to
the functional $\tau_{\infty}$
$$
J_{\infty} = {D_0 P_0 \over \tau_{\infty} }
\eqno (10)
$$
Note that unlike the velocity $V$, this flux is  not
a self-averaging quantity,  but must be described
by its full probability distribution.

It is very convenient to introduce the more general functional
$$
\tau(x,b) = \int_x^b dy \ e^{ \displaystyle \beta \left[U(y)-U(x) \right ]}
 = \int_x^b dy \ e^{ \displaystyle - \beta \int_x^y F(u) du}
\eqno (11)
$$
and to consider the random variable $\tau_{\infty}$ as
the limit of this process as $x \to -\infty$
$$
\tau_{\infty} = \lim_{x \to -\infty} \ \tau(x,b)
\eqno (12)
$$
The evolution of the functional $\tau(x,b)$ is governed by the the
stochastic differential equation
$$
{{\partial \tau } \over {\partial x }} = \beta F(x) \tau(x,b) -1
\eqno (13)
$$
The stochastic term $F(x)$ appears multiplicatively, so that the
fluctuations of the random force are coupled to the values taken by the random
process
$\tau(x,b)$ itself. When the random force $\{F\}$ is distributed as a white
noise (8), the
multiplicative stochastic process $\tau(x,b)$ can be in fact be related to
Brownian motion on a surface of constant negative curvature [9].

We now compute the probability distribution of the functional $\tau_{\infty}$
when the quenched random force $\{F(x)\}$ is a two-level Poisson process,
and compare it with the result (9) for the white noise case.
The random variable $\tau_{\infty}$ only exists if the force mean value $F_0$
is strictly positive
$$
(n_0+n_1)F_0 = \phi_0n_1-\phi_1n_0 >0 \quad \quad \hbox{that is} \quad
{n_1 \over \phi_1}-{n_0 \over \phi_0}>0
\eqno(14)
$$
that we assume from now on.
Note that the random variable $\tau(x,b)$ remains confined in
an interval $[\tau _{min} (x,b),\tau _{max} (x,b)]$
depending on the length  $(b-x)$
 $$
\left\{
\matrix{
&\tau _{min} (x,b) =\int_x^b dy \ e^{-\beta \phi_0 (y-x)} = \displaystyle
{     {1- e^{-\beta \phi _0 ( b-x )}} \over {\beta \phi _0}         } \cr
\hbox{} \cr
&\tau _{max} (x,b) =\int_x^b dy \ e^{+\beta \phi_1 (y-x)} = \displaystyle
{{ e^{\beta \phi _1 ( b-x )} -1} \over {\beta \phi _1}} \hfill \cr
} \right\}
\eqno (15)
$$
This is very different from the white noise case, where the random force is not
bounded.
Even in the limit where the interval length $(b-x)$ tends to infinity,
the support of the random variable $\tau_{\infty}$
is not $[0, + \infty]$, but $[{1 \over {\beta \phi _0}}, + \infty]$.

The two-level Poisson process $\{F\}$ is Markovian; so according to the local
evolution equation (13), the coupled process
 $\{F(x), \tau(x)\}$ is still Markovian.
Let us define the joint laws
$$
P_0(\tau,x) d\tau = \hbox{Prob} \bigg\{ F(x) = \phi_0 \ \hbox{and}
\ \tau (x,b) \in [\tau, \tau+d\tau] \bigg\}
$$
$$
P_1(\tau,x) d\tau = \hbox{Prob} \bigg\{ F(x) = -\phi_1 \ \hbox{and}
\ \tau (x,b) \in [\tau, \tau+d\tau] \bigg\}
\eqno (16)
$$
They evolve according to the two coupled Master equations
$$
\left\{
\matrix{
\displaystyle - {\partial P_0 \over \partial x} &= - \displaystyle
{\partial  \over \partial \tau }
 \big[ (1-\beta \phi _0 \tau) P_0 \big] -n_0 P_0 + n_1 P_1 \cr
\hbox{} \cr
\displaystyle - {\partial P_1 \over \partial x} &= - \displaystyle
{\partial  \over \partial \tau }
\big[ (1+\beta \phi _1 \tau) P_1 \big] +n_0 P_0 - n_1 P_1 \cr
}\right\}
\eqno (17)
$$
Stationary distributions $P_0(\tau)$ et $P_1(\tau)$ in the limit $x \to
-\infty$
are therefore the solutions of the system
$$
\left\{
\matrix{
\displaystyle {d  \over d \tau }  \big[ (\beta \phi _0 \tau -1) P_0 \big] -n_0
P_0 + n_1 P_1 =0 \cr
\noalign{\medskip}
\displaystyle {d  \over d \tau }  \big[ (\beta \phi _1\tau +1) P_1 \big] - n_0
P_0 + n_1 P_1 =0\cr
}\right\}
\eqno (18)
$$
respectively normalized on the interval $[{1 \over {\beta \phi _0}}, + \infty]$
by
$$
\int_{\displaystyle {1 \over {\beta \phi _0}}}^{\infty} P_0(\tau) d \tau = {n_1
\over {n_0 +n_1}}
\ \ \ \hbox{and} \ \ \
\int_{ \displaystyle {1 \over {\beta \phi _0}}}^{\infty} P_1(\tau) d \tau =
{n_0 \over {n_0 +n_1}}
\eqno (19)
$$

It is convenient to set $\tau _0 = \displaystyle {1\over {\beta \phi _0}}$ ,
  $\tau _1 =  \displaystyle {1 \over {\beta \phi _1}}$ , $\nu_0 =n_0 \tau_0$ et
$\nu_1 =n_1 \tau_1$,
and
$$
\nu = \nu_1 - \nu_0 ={1 \over \beta} \left(   {n_1 \over \phi_1} -{n_0 \over
\phi_0} \right) >0
\eqno (20)
$$
which is strictly positive according to the hypothesis $F_0>0$ (14).
The solutions of (18) (19) read, using Heaviside function $\theta$ ,
$$
\left\{
\matrix{
P_0(\tau) =A \ \theta (\tau - \tau_0) \displaystyle {{(\tau - \tau_0)^{\nu_0
-1}} \over
{ (\tau + \tau _1)^{ \nu_1}}}
\ &\hbox{where}   &A={n_1 \over {n_0 +n_1}} { {\Gamma(\nu_1 )} \over
{\Gamma(\nu _0)
\Gamma (\nu_1 - \nu_0)}} (\tau_0 + \tau _1)^{\nu_1 - \nu_0} \cr
\noalign{\medskip}
P_1(\tau) =B \ \theta (\tau - \tau_0) \displaystyle {{(\tau - \tau_0)^{\nu_0 }}
\over
{ (\tau + \tau _1)^{ \nu _1 +1}}}
&\hbox{where}  &B={n_0 \over {n_0 +n_1}} { {\Gamma(\nu_1 +1)} \over {\Gamma(\nu
_0 +1)
\Gamma (\nu_1 -\nu_0)}}
(\tau_0 + \tau _1)^{\nu_1 -\nu_0} \cr
}\right\}
\eqno (21)
$$
Moments of order $k$ of these distributions diverge as soon as
$k \geq \nu=(\nu_1-\nu_0)$. They read otherwise in polynomial forms as
$$
\left\{
\matrix{
\int_0^{\infty} d\tau \ \tau^k \ P_0(\tau) &=
{ \displaystyle {n_1 \over {n_0 +n_1}}} \  &\tau_0^k \  &F \left(-k, \nu_0,
1+\nu_0-\nu_1;
 1+\displaystyle{\tau_1 \over \tau_0}\right)\cr
\noalign{\medskip}
\int_0^{\infty} d\tau \ \tau^k \ P_1(\tau) &=
{ \displaystyle{n_0 \over {n_0 +n_1}}} &\tau_0^k &F \left(-k, \nu_0 +1,
1+\nu_0-\nu_1;
 1+\displaystyle{\tau_1 \over \tau_0}\right)\cr
}\right\}
\eqno (22)
$$
The velocity can be directly deduced from first moment of $\tau_{\infty}$ (7)
$$
{1 \over V} = {1 \over D_0} < \tau _{\infty} > ={1 \over D_0}
 \int_0^{\infty} d \tau  \ \tau \ \big[ P_0(\tau) + P_1(\tau) \big]
\eqno (23)
 $$
There exists therefore a dynamical phase transition at  $\nu = (\nu_1-\nu_0) =
1$
$$
V= \left\{ \matrix{
&0 \hfill\ \ &\hbox{if} \ \ &\nu \leq 1 \cr
&\hbox{} \cr
&\displaystyle{ {D_0 \left(\nu-1\right)} \over { \displaystyle
{n_1 \over {n_0 +n_1}} \left[ \tau_0 (\nu_1 -1)
+\tau_1 \nu_0\right]+ {n_0 \over {n_0 +n_1}} \left[ \tau_0 \nu_1  +\tau_1
(\nu_0+1) \right]}}
\ \ &\hbox{if} \ \ &\nu \geq 1 \cr
}\right\}
\eqno (24)
$$
More generally, the algebraic decay of the distributions $P_{0,1}(\tau) $ in
the limit
$\tau \to \infty$
$$
P_{0,1}(\tau) \oppropto_{\tau \to \infty} {1 \over {\tau}^{1+ \nu}}
\ \ \ \ \ \hbox{with} \ \ \ \ \
\nu = \nu_1 - \nu_0 ={1 \over \beta} \left(   {n_1 \over \phi_1} -{n_0 \over
\phi_0} \right)
\eqno (25)
$$
will generate the same succession of anomalous diffusion behaviours for
the cumulants of the Brownian particle position as a function of the parameter
$\nu$,
as the one that appears in the white noise case
 as a function of the parameter $\mu$ [3]. In particular, the thermal average
of the position
 of the particle will grow linearly in time only in the finite velocity phase
 $$
 \overline {x(t)} \equiv \int_{- \infty}^{+ \infty} dx \ x \ P(x,t \  \vert
x_0, 0)
 \oppropto_{t \to \infty} \ V \ t  \ \ \ \ \ \hbox{if} \ \  \nu > 1
 \eqno (26)
 $$
However, in the vanishing velocity phase, the thermal average of
the position will grow slower than linearly,
as a power of time with the exponant $\nu$ which depends
continuously on the parameters of the disorder
 $$
 \overline {x(t)} \equiv \int_{- \infty}^{+ \infty} dx \ x \ P(x,t \  \vert
x_0, 0)
 \oppropto_{t \to \infty} \ \ t^{\nu} \ \ \ \ \ \ \hbox{if} \ \  0<\nu < 1
 \eqno (27)
 $$

Let us now consider the stationary flux $J_{\infty}$
 going through a sample between two concentrations $P_0>0$ and $P_{N}=0$
 in the limit $N \to \infty$.
 The change of variable (10) gives immediately the two joint
laws ${\cal P}_0(J)$ et ${\cal P}_1(J)$ from result (21), with the notations
$J_0 ={{D_0 P_0} \over {\tau _0}}$ and $J_1 = {{D_0 P_0} \over {\tau _1}}$
$$
\left\{
\matrix{
{\cal P}_0(J) ={\cal A} \ \theta (J) \ \theta (J_0 -J) \ J^{\nu_1-\nu_0 -1} \
\displaystyle {{(J_0 -J)^{\nu_0 -1}} \over { (J_1 + J)^{ \nu_1}}}  \cr
\noalign{\medskip}
{\cal P}_1(J) = {\cal B} \ \theta (J) \ \theta (J_0 -J) \ J^{\nu_1-\nu_0 -1} \
\displaystyle {{(J_0 -J)^{\nu_0 }} \over { (J_1 + J)^{ \nu_1 +1}}}   \cr
}\right\}
\eqno (28)
$$
where ${\cal A}$ and ${\cal B}$ are two normalization constants. \hfill \break
These two distributions for the flux are concentrated on the bounded interval
$[0, J_0]$.
The transition at $\nu \equiv \nu_1 - \nu_0=1$ for the self-averaging velocity
(24)
corresponds for the flux probability distributions to a transition of the
behaviour
in the limit ${J \to 0}$
$$
{\cal P}_{0,1} (J) \ \oppropto_{J \to 0} \ J^{\nu-1}  \longrightarrow \left\{
\matrix{
&+ \infty \hfill\ \ &\hbox{if} \ \ &\nu < 1 \cr
&\hbox{} \cr
&0  \ &\hbox{if} \ \ &\nu > 1 \cr
}\right\}
\eqno (29)
$$
The vanishing velocity phase $\nu =\nu_1 -\nu_0<1$ is characterised
by the divergence of ${\cal P}_0$ and ${\cal P}_1$ at the origin $J \to 0$.
 On the contrary in the finite velocity phase $\nu>1$, the probability
distributions
 of the flux vanish at the origin.

Some curves ${\cal P}_{0,1} (J)$ are drawn on Figures 2 and 3 for different
values of
the parameters. They illustrate in particular the transitions at $\nu=1$
we just mentioned (29) and the transion at $\nu_0=1$ for the behaviour of
${\cal P}_0(J)$
as $J \to J_0$. We refer to the figures in [8] for comparison with the
uncorrelated case.

In this article, we have analysed some properties of anomalous diffusion in a
random
medium described by a quenched random force $\{F\}$
distributed as a two-level Poisson process.
In particular, we showed that there exists a dimensionless parameter $\nu$,
which governs
the asymptotic behaviours of the probability distributions
 $P_{0,1} (\tau)$ in the limit $\tau \to \infty$. As a result, this parameter
$\nu$
 also controlls the dynamical phase transition for the velocity and the
behaviours
of the probability distributions
 ${\cal P}_{0,1}(J)$ of the flux in the limit $J \to 0$. All this is
qualitatively
the same as what is known in the white noise case.
However, the probability distributions for the random variable $\tau_{\infty}$
and for the stationary flux $J_{\infty}$ are very different from the white
noise case
outside the asymptotic regimes $\tau \to \infty$ and $J \to 0$. Indeed, they
present
restricted supports, as a consequence of the bounded character of the random
force,
and they are given in terms of only rational functions and no exponential.

More generally, the two-level Poisson process we considered is a technically
very convenient disorder model, since it allows analytical studies despite
the presence of correlations. It has already been used
in various contexts, like one-dimensional quantum localization [10] [11],
non-linear systems coupled to a random environment [12], and noise-induced
 perturbations on Josephson junctions [13]. The limitation to two-level
processes simplifies computations, but the approach can be generalized to any
finite number of levels for the random process [14].

\vskip 1 true cm

\leftline {\bf Acknowledgements}

\vskip 0.5 true cm

I wish to thank Alain Comtet for many helpful discussions and for his remarks
on
the manuscript.

\vfill \eject

\vskip 1 true cm

\leftline {\bf References}

\vskip 0.5 true cm

\item{\hbox to\parindent{\enskip [1] \hfill}}
F. Solomon, {\it Ann. Prob.} {\bf 3} (1975) 1;
H. Kesten, M. Koslov and F. Spitzer, {\it Compositio Math. } {\bf 30 } (1975)
145;
Y.A.G. Sina\"{\i}, {\it Theor. Prob. Appl.} {\bf XXVII} (1982) 256;
B. Derrida and Y. Pomeau, {\it Phys. Rev. Lett. } {\bf 48 } (1982) 627;
B. Derrida, {\it J. Stat. Phys.} {\bf 31} (1983) 433;

\item{\hbox to\parindent{\enskip [2] \hfill}}
 J.W. Haus and K.W. Kehr, {\it Phys. Rep.} {\bf 150} (1987) 263;
S. Havlin and D. Ben Avraham,  {\it Adv. Phys.} {\bf 36} (1987) 695.

\item{\hbox to\parindent{\enskip [3] \hfill}}
J.P. Bouchaud and A. Georges, {\it Phys. Rep.} {\bf 195} (1990) 127;
J.P. Bouchaud, A. Comtet, A. Georges and P. Le Doussal,
 {\it Ann. Phys. } {\bf 201} (1990) 285;
A. Georges, th\`ese d'\'etat de l'Universit\'e Paris 11 (1988).

\item{\hbox to\parindent{\enskip [4] \hfill}}
C. Aslangul, N. Potier, P. Chvosta and D. Saint-James,
{\it Phys. Rev. E} {\bf 47} (1993)
1610.

\item{\hbox to\parindent{\enskip [5] \hfill}}
P. Le Doussal and V. Vinokur, preprint cond-mat / 9501131 ;
S. Scheidl, preprint cond-mat / 9501121.

\item{\hbox to\parindent{\enskip [6] \hfill}}
C. Aslangul, N. Potier and D. Saint-James,  {\it J. Phys. (France)} {\bf 50}
(1989), 899.

\item{\hbox to\parindent{\enskip [7] \hfill}}
S.F. Burlatsky, G.H. Oshanin, A.V. Mogutov and M. Moreau,
 {\it Phys. Rev. A} {\bf 45 } (1992) 6955;
G. Oshanin, A. Mogutov and M. Moreau, {\it J. Stat. Phys.} {\bf 73 } (1993);
G. Oshanin, S.F. Burlatsky, M. Moreau and B. Gaveau, {\it Chem. Phys.} {\bf 178
} (1993).

\item{\hbox to\parindent{\enskip [8] \hfill}}
C. Monthus and A. Comtet, {\it J. Phys. I (France)} {\bf 4} (1994) 635.

\item{\hbox to\parindent{\enskip [9] \hfill}}
A. Comtet and C. Monthus, preprint.

\item{\hbox to\parindent{\enskip [10] \hfill}}
M.M. Benderskii and L.A. Pastur, {\it Sov. Phys. JETP} {\bf 30 } (1970) 158;
L.A. Pastur and E.P. Fel'dman, {\it Sov. Phys. JETP} {\bf 40} (1974) 241;
I.M. Lifshits, S.A. Gredeskul and L.A. Pastur,
``Introduction to the theory of disordered systems",
John Wiley and Sons, New York (1987) .

\item{\hbox to\parindent{\enskip [11] \hfill}}
A. Comtet, J. Desbois and C. Monthus, {\it Ann. Phys.} { \bf 239} (1995) 312.

\item{\hbox to\parindent{\enskip [12] \hfill}}
W. Horsthemke and R. Lefever, ``Noise-induced transitions : theory and
applications in physics,
 chemistry and biology",
 Springer Verlag (1984).

\item{\hbox to\parindent{\enskip [13] \hfill}}
G.H. Weiss and M. Gitterman, {\it J. Stat. Phys.} {\bf 70 } (1993) 93.

\item{\hbox to\parindent{\enskip [14] \hfill}}
P. Erd\"os and Z. Domanski, in ``Analogies in optics and microelectronics"
(W. Van Haeringen and
D. Lenstra eds.) Kluwer Academic Publishers (1990) 49.

\vfill \eject

\vskip 1 true cm

\leftline {\bf Figures captions}

\vskip 0.5 true cm

\item{\hbox to\parindent{\enskip 1) \hfill}}
 Example of the random potential $U(x)$ seen by the Brownian particle
when the random force $\{F\}$ is a two-level Poisson process.

\item{\hbox to\parindent{\enskip 2) \hfill}}
Examples of flux probability distributions ${\cal P}_0(J)$ (plain line)
and ${\cal P}_1(J)$ (dashed line) in the vanishing velocity phase $0<\nu<1$
for \hfill \break
2a) $\nu_0 <1$ \hfill \break
2b) $\nu_0 >1$

\item{\hbox to\parindent{\enskip 3) \hfill}}
Examples of flux probability distributions ${\cal P}_0(J)$ (plain line)
and ${\cal P}_1(J)$ (dashed line) in the finite velocity phase $\nu>1$
for \hfill \break
3a) $\nu_0 <1$ \hfill \break
3b) $\nu_0 >1$

\end